\documentclass[lettersize,journal]{IEEEtran}
\usepackage{amsmath,amsfonts}
\usepackage{amsmath,amssymb,amsfonts}
\usepackage{amsmath,amsthm,amssymb,mathtools}
\usepackage{algorithmic}
\usepackage{algorithm}
\usepackage{array}
\usepackage{acronym}
\usepackage[caption=false,font=normalsize,labelfont=sf,textfont=sf]{subfig}
\usepackage{textcomp}
\usepackage{stfloats}
\usepackage{url}
\usepackage{verbatim}
\usepackage{graphicx}
\usepackage{cite}
\usepackage{xcolor}%
\usepackage{todonotes}
\newacro{QNT}{Quantum Network Tomography}
\newacro{QFI}{Quantum Fisher Information}
\newacro{CFI}{Classical Fisher Information}
\newacro{CFIM}{Classical Fisher Information Matrix}
\newacro{QFIM}{Quantum Fisher Information Matrix}
\newacro{FIM}{Fisher Information Matrix}
\newacro{QCRB}{Quantum Cramér-Rao Bound}
\newacro{CRB}{Cramér-Rao Bound}
\newacro{MLEs}{Maximum Likelihood Estimators}
\newacro{MLE}{Maximum Likelihood Estimator}
\newacro{BSMs}{Bell-state measurements}
\newacro{BSM}{Bell-state measurement}
\newacro{SLDs}{Symmetric Logarithmic Derivatives}
\newacro{SPAM}{State Preparation and Measurement Errors}

\begin{document}
\title{Identifiability and Estimation Precision in Quantum Network Tomography with Imperfect Bell-State Measurements
}

\author{\IEEEauthorblockN{Athira Kalavampara Raghunadhan\textsuperscript{1},  Matheus Guedes De Andrade\textsuperscript{2}, Don Towsley\textsuperscript{2}, Indrakshi Dey\textsuperscript{3},\\   Daniel Kilper\textsuperscript{1}, Nicola Marchetti\textsuperscript{1} }
\\
\IEEEauthorblockA{\textsuperscript{1}CONNECT Research Centre, School of Engineering, Trinity College Dublin, Ireland}
\\
\IEEEauthorblockA{\textsuperscript{2}Manning College of Information and Computer Science, University of Massachusetts, Amherst, USA}
\\
\IEEEauthorblockA{\textsuperscript{3}Department of Computing, School of Science, South East Technological University, Waterford, Ireland}}


\markboth{Journal of \LaTeX\ Class Files,~Vol.~14, No.~8, August~2021}%
{Shell \MakeLowercase{\textit{et al.}}: A Sample Article Using IEEEtran.cls for IEEE Journals}



\maketitle

\begin{abstract}
We study Quantum Network Tomography (QNT) for end-to-end link-error characterization under imperfect Bell-state measurements (BSMs), where multiplicative coupling between link and measurement parameters makes identifiability non-trivial. For an $n$-node star network, we design probes that ensure unique identifiability and derive closed-form expressions for the Fisher Information Matrix (FIM) and Maximum Likelihood Estimators (MLEs), and characterize estimation precision through the Cramér–Rao Bound (CRB). The results show that BSM imperfections degrade estimation precision, while the proposed probes maintain nearly stable precision for individual link-parameters as the network size increases. Monte Carlo simulations further confirm that the Mean Squared Error (MSE) approaches the CRB with increasing sample size.

\end{abstract}


\section{Introduction}

\IEEEPARstart{Q}{uantum} communication requires accurate characterization of quantum links and operations, as their imperfections directly impact network performance. \ac{QNT} \cite{de2022quantum, de2024quantum, de2023characterization} aims to estimate unknown parameters of quantum links using carefully designed probe states and end-to-end measurements performed at selected network nodes, called monitors \cite{he2021network}. 

In QNT, the entangled probe states are distributed through the network along selected paths. As these probe states traverse the quantum channels, they encode information about the error parameters of the links forming those paths. The monitors then perform measurements on the received probe states, and the resulting measurement statistics are used to infer the underlying link parameters, while regular network nodes are assumed to perform only the \ac{BSMs} required for communication.

Most existing works \cite{optimalqnt,raghunadhan2026optimization} on QNT assume an ideal setting, where operations such as entanglement distribution and measurements are perfect, and only the link parameters need to be estimated. Under these assumptions, probe state distribution and measurement strategies have been proposed, and their estimation performance has been evaluated using the \ac{QFIM} and the \ac{QCRB} \cite{liu2020quantum}. These studies have provided useful insights into monitor placement and measurement assignment in arbitrary network topologies under ideal conditions.

Recent work \cite{wang2025quantum} presents an important step toward realistic \ac{QNT} by accounting for \ac{SPAM} at peripheral monitors. However, entanglement swapping relies on intermediate \ac{BSMs}, which are themselves subject to operational imperfections. Their unknown noise parameters couple with the unknown link parameters, creating a distinct joint-identifiability problem and degrading the achievable estimation precision. In this work, we address this gap by jointly characterizing link and \ac{BSM} parameters in star networks under the assumption that both link-level entanglement and BSMs are corrupted by depolarizing noise. The main contributions are summarized as follows:
\begin{itemize}
        \item We show that neglecting BSM imperfections produces a systematic underestimate of the link parameters and characterize the resulting identifiability problem caused by multiplicative coupling between the link and BSM depolarizing parameters.
        \item We show that probe design plays a critical role in ensuring parameter identifiability under imperfect \ac{BSMs}, and demonstrate that introducing a locally generated entanglement resource in probe design restores unique identification of all parameters.
        \item We derive closed-form expressions for the \ac{FIM} and \ac{MLEs} for an $n$-node star network and evaluate the estimation accuracy using the \ac{CRB}.  
\end{itemize}
The remainder of the paper is structured as follows. Section II defines the system model, Section III reports the numerical evaluation results, and Section IV concludes with a summary of the main findings and future directions.

\section{System model} 
We model the quantum network as a graph $G=(V,E)$, where $V$ denotes the set of quantum processors and $E$ denotes the set of quantum links. Each link $e_k \in E$ is modeled as a depolarizing channel. Neighboring nodes generate link-level entanglement through these noisy channels, producing Werner states \cite{sen2005entanglement} of the form
\begin{align}
    \rho(w_k) = w_k \Phi^{+} + \left(1 - w_k\right) \frac{I}{4},
\end{align}
where $w_k \in (0, 1)$ characterizes the entanglement quality of link $e_k$. In this paper, we focus on an $n$-node star network with central node $v_0$ and leaf nodes $v_1,\ldots,v_{n-1}$. The network has $L = n - 1$ links, where $e_k=(v_0,v_{k+1})$ for $k=0,\ldots, L - 1$. Throughout the paper, $e_u$ and $e_v$ denote two distinct links in the star network, and $\rho(w)$ denotes a Werner state with parameter $w \in (0, 1)$. Entanglement between non-neighboring nodes is generated through \ac{BSMs}, i.e., entanglement swapping, at the intermediate node. \ac{BSM}s are assumed to be imperfect and modeled as ideal \ac{BSM}s followed by a depolarizing channel, assuming the form $\mathcal{B}_{b}=(I \otimes \mathcal{D}_{b})\circ \mathcal{B}$, where $\mathcal{B} : \mathcal{H}^{16} \to \mathcal{H}^{4}$ denotes the ideal \ac{BSM} channel, $I$ denotes the one-qubit identity operator, and
\begin{equation}
    \mathcal{D}_{b}(\sigma) = b \sigma + (1 - b)\frac{I}{2}
\end{equation}
is the one-qubit depolarizing channel with parameter $b \in (0, 1)$, and $\sigma: \mathcal{H}^2\to \mathcal{H}^2$ is a one-qubit density operator. Thus, a probe involving $m$ \ac{BSMs} contains the factor $b^m$. The unknown parameter vector is 
\mbox{$\boldsymbol{\theta} = (w_0,\ldots,w_{L-1},b) \in (0, 1)^{L + 1}$}, containing the Werner parameters of all links and the \ac{BSM} depolarizing parameter.
For networks with multiple intermediate nodes, the identification of node-specific BSM parameters would require additional probe configurations beyond those considered in this work.
\vspace{-1.0em}
\subsection{Probe State Construction}
\begin{figure}[t]
\centering
\subfloat[\label{p1}]{\includegraphics[scale=0.22]{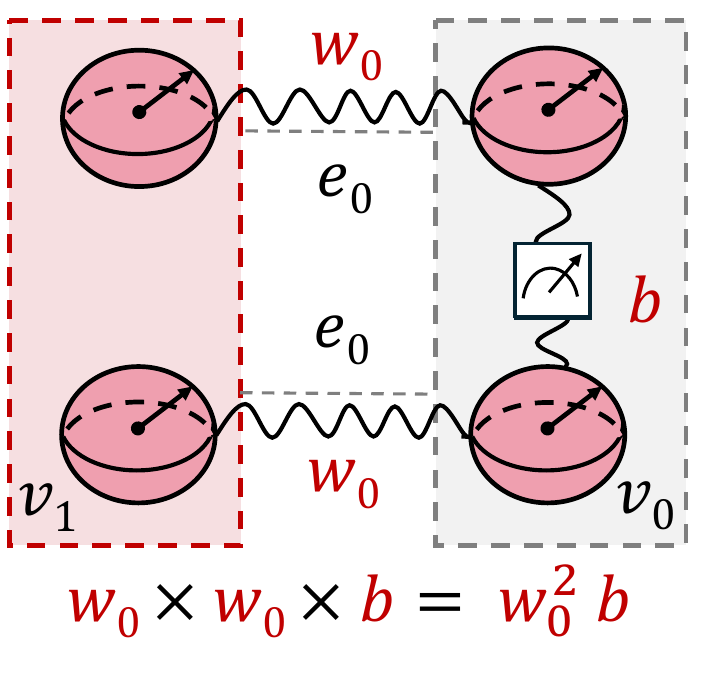}}%
\hspace{1.5mm}
\subfloat[\label{p2}]{\includegraphics[scale=0.22]{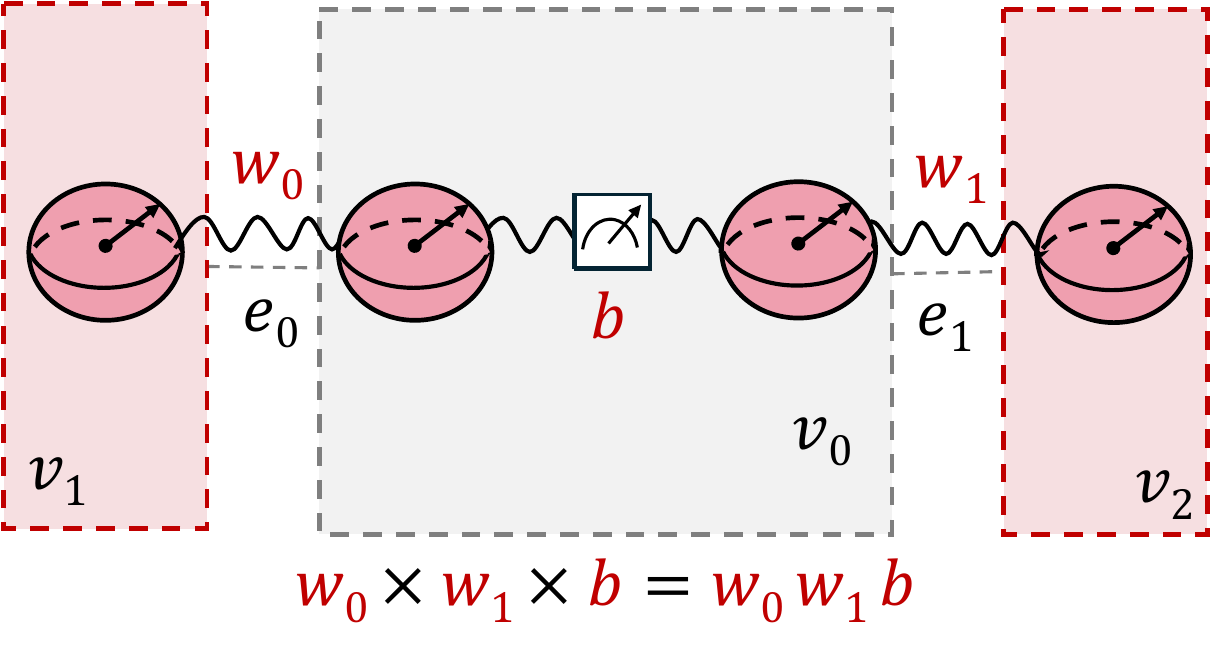}}%
\vspace{-0.5em}
\caption{Probe construction in a three-node network with monitors at $v_1$ and $v_2$. Links $e_0=(v_1,v_0)$ and $e_1=(v_0,v_2)$ are characterized by $w_0$ and $w_1$, respectively, and each BSM at the intermediate node $v_0$ is characterized by $b$: (a) cyclic probe over $e_0$, yielding $w_0^2b$, (b) linear probe, yielding $w_0w_1b$.\label{fig1}}
\end{figure}

In \ac{QNT}, the probing strategy determines how channel parameters are encoded in the observed measurement outcomes. In particular, probe states generated over individual links provide direct information about the corresponding link parameter, while probe states generated over multi-link paths encode products of the parameters associated with the traversed links. Following prior work~\cite{optimalqnt}, we use two probe-state generation methods for parameter estimation, namely \emph{linear} and \emph{cyclic} generation, with local $Z\otimes Z$ and Bell-basis measurements, respectively. In linear generation, entanglement is distributed along a path between two monitors, with intermediate nodes performing \ac{BSMs} to generate an end-to-end state. In cyclic generation, two end-to-end states are combined through a \ac{BSM} at one end node, so that only one monitor is required on the path. As a result, linear generation through path $(e_u, e_v)$ produces the Werner state $\rho(w_u w_v b)$ with effective end-to-end parameter $w_u w_v b$, which depends linearly on the link parameter, while cyclic generation through link $e_k$ produces the Werner state $\rho(w_k^2 b)$, which depends quadratically on the link parameter as shown in Fig. \ref{fig1}.

The way a probe state is generated determines how each link is observed in the tomography process. Based on this, we classify the monitoring process as \textit{direct} when a monitor estimates a link $e_i$ using an entangled state generated solely over $e_i$. In contrast, when the entangled state is distributed over a multi-link path that includes $e_i$, the link is considered to be \textit{indirectly monitored}.
\vspace{-1.0em}
\subsection{Bias under Ideal BSM}
BSM imperfections result in biased link estimates. To illustrate this, consider the cyclic probe associated with link $e_k$, the Bell-basis outcome probabilities are
\begin{equation}
p_{\Phi^+}
=
\frac{1+3w_k^2b}{4},
\qquad
p_{\Phi^-}=p_{\Psi^+}=p_{\Psi^-}
=
\frac{1-w_k^2b}{4}.
\label{eq:single_link_prob}
\end{equation}
If the BSM is incorrectly assumed to be ideal, i.e., $b=1$, these measurement statistics are interpreted as arising from a link parameter $\widetilde{w}_k$ satisfying $\widetilde{w}_k^2=w_k^2b$ and $\widetilde{w}_k=w_k\sqrt{b}$. Consequently, an ideal BSM estimator $\widehat{w}_{k,\mathrm{ideal}}$ applied to imperfect BSM data converges to $w_k\sqrt{b}$ as the number of measurement samples $N \rightarrow \infty$, with asymptotic bias
\mbox{$
w_k(\sqrt{b}-1)$}.
Thus, for $b<1$, the link parameter $w_k$ is systematically underestimated as $\widetilde{w}_k$, and this error does not vanish as the number of measurements increases. This motivates the joint identification of the link and BSM parameters rather than treating the BSM as ideal.
\vspace{-1.0em}
\subsection{Identifiability Problem}
We now examine whether the link parameters and the imperfect-BSM parameter can be uniquely identified in an $n$-node star network. The parameter vector \mbox{$\boldsymbol{\theta}$} is identifiable if the mapping from $\boldsymbol{\theta}$ to the outcome distributions of the probe set is injective. We use the probe-state generation approaches of Ref.\cite{optimalqnt}, which were originally studied under ideal \ac{BSM} conditions. 

For each link $e_k$, the cyclic probe as shown in Fig.~\ref{p1} generates the Werner state $\rho(w_k^2 b)$ whose effective parameter is $x_k=w_k^2b$, for $k=0,\ldots,L - 1$, and then the corresponding monitor performs a \ac{BSM} to obtain classical outcomes for parameter estimation. In addition, we consider a linear probe over the two-link path involving $e_u$ and $e_v$ which generates the end-to-end Werner state $\rho(w_u w_v b)$ with effective parameter $z_{uv}=w_uw_vb$ as shown in Fig.~\ref{p2}. The corresponding monitor nodes then perform local measurements in the computational basis ($Z$) on their respective qubits to estimate the parameters. Thus, the observed probe states contain the multiplicative combinations of the unknown parameters given by
\mbox{$
\left\{
w_k^2b:k=0,\ldots,L - 1
\right\}
\cup
\left\{
w_uw_vb
\right\}.
$}

To study identifiability, we convert these combinations into linear form by taking logarithms. The boundary cases $w_k=0$ and $b=0$ are excluded, ensuring that the logarithmic transformation is well defined. Let
$
y_k=2\log w_k+\log b$, for $k=0,\ldots,L - 1,
$ and
$
y_L=\log w_u+\log w_v+\log b.
$
This can be written in matrix form as $\mathbf{y}=A \mathbf{x}$, where
$\mathbf{y}=(y_0,\ldots,y_L)^T,\,
\mathbf{x}=(\log w_0,\ldots,\log w_{L - 1},\log b)^T,
$
and the $(L+1)\times(L+1)$ matrix $A$ is
\begin{equation}
    A = \begin{bmatrix}
        2I_L & \mathbf{1}_L\\
        \mathbf{e}_u^T+\mathbf{e}_v^T & 1
        \end{bmatrix}.
\end{equation}
Here, $I_L$ is the $L\times L$ identity matrix, $\mathbf{1}_L$ is the all-ones column vector of length $L$, and $\mathbf{e}_u$ denotes the $u$-th canonical basis vector.  Since the logarithmic transformation is one-to-one on the considered domain, identifiability holds if, and only if, $A$ has full column rank.  However, the last row is linearly dependent on the rows corresponding to links $e_u$ and $e_v$. 

Let $R_k$ denote the row associated with the single-link probe $\rho(w_k^2b)$, for \mbox{$k=0,\ldots, L-1$}, and $R_{\mathrm{p}}$ denote the row associated with the two-link probe $\rho(w_uw_vb)$. Then
$
\label{rpair}
R_{\mathrm{p}}=\frac{1}{2}R_u+\frac{1}{2}R_v.
$
Since the first $L$ rows of $A$ are linearly independent, $\mathrm{rank}(A) \geq L$. From $R_p$, the last row does not add any new information. Hence, $\mathrm{rank}(A)=L<L+1$. Therefore, $A$ does not have full column rank, and the parameters cannot be uniquely identified from these probe observations. This motivates the need for additional probing capability to achieve unique identifiability.

\vspace{-1.0em}
\subsection{Linear probe construction with additional entangled pair}
\begin{figure}[t]
\centering
\includegraphics[scale=0.22]{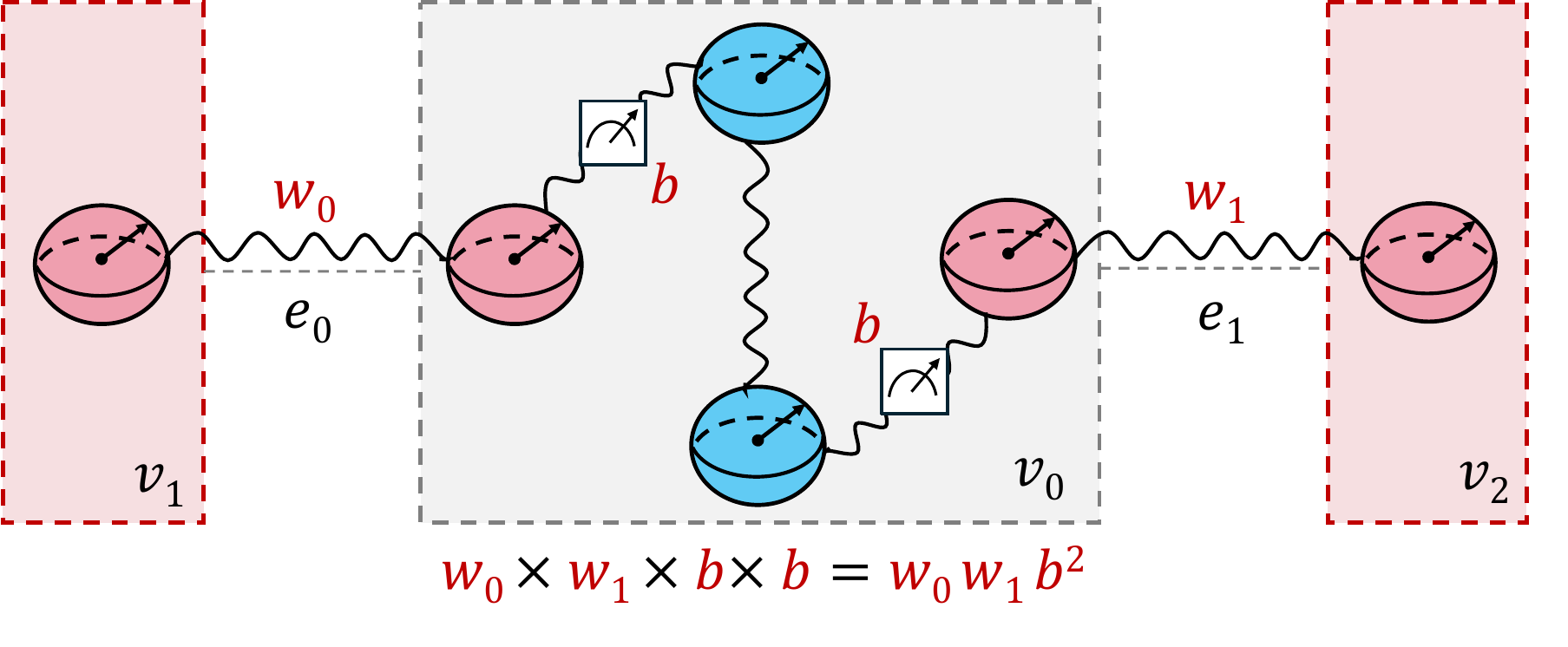}%
\vspace{-1.0em}
\caption{Two-link linear probe with a locally generated entangled pair at $v_0$, yielding $w_0w_1b^2$.}
\label{p3}
\end{figure}
To address the identifiability limitation observed above, we introduce an additional capability in linear probe construction. We assume that the central node in the star network has access to a locally available perfect entangled pair. This additional resource allows us to construct probe states that isolate different noise contributions more effectively. In particular, locally available entanglement can be combined with distributed entangled states to create end-to-end entangled states that encode parameters differently and yield parameter identifiability.

We now construct probes by incorporating locally available entanglement. The cyclic probes of the form $\rho(w_k^2 b)$ remain unchanged as shown in Fig.~\ref{p1}. The linear probe involving the two-link path formed by $e_u$ and $e_v$ is altered to include two \ac{BSMs} at the central node $v_0$, as shown in Fig.~\ref{p3}. Since these operations are carried out locally at the same node, we assume they experience identical noise captured by a depolarizing channel with parameter $b$. Hence their combined contribution to the effective noise parameter is $b^2$. As a result, the corresponding end-to-end Werner state is given by $\rho(w_u w_v b^2)$. Thus, the complete set of effective parameter combinations used for estimation is
\mbox{$
\label{eq:effective_parameter_set}
\left\{
w_k^2b:k=0,\ldots,L - 1
\right\}
\cup
\left\{
w_uw_vb^2
\right\}.
$}

For the case $(u,v)=(0,1)$, the first two cyclic probes provide
$w_0^2b$ and $w_1^2b$, while the additional two-link probe provides $w_0w_1b^2$. These three combinations are sufficient to identify $w_0$, $w_1$, and $b$. Once $b$ is identified, each remaining link parameter $w_k$ can be recovered from the corresponding combination $w_k^2b$, for \mbox{$k=2,\ldots, L - 1$}. Taking logarithms of the effective parameter set, we obtain
\mbox{$y_k=2\log w_k+\log b$}, for \mbox{$k=0,\ldots,L-1$}, and
$y_L=\log w_u+\log w_v+2\log b$. The corresponding linear system is
$
\mathbf{y}=A'\mathbf{x},
$
where
\begin{equation}
A'=
\begin{bmatrix}
2I_L & \mathbf{1}_L\\
\mathbf{e}_u^T+\mathbf{e}_v^T & 2
\end{bmatrix}.
\end{equation}
The parameters are identifiable when $A'^{-1}$ exists, \textit{i.e.}, $\det(A')\neq 0$.
Let $R_k$, for \mbox{$k=0,\ldots, L - 1$}, and $R_p$ denote the first $L$ rows and the last row of $A'$, respectively.
Applying the row operation
\mbox{$R_p \leftarrow R_p-\frac{1}{2}R_u-\frac{1}{2}R_v$}
reduces the last row to $[0,\ldots,0,1]$.
Since $\det(A') = 2^L \neq 0$, $A'$ is invertible and $\operatorname{rank}(A')=L + 1$.
Therefore, the parameters $\{w_0,\ldots,w_{L - 1},b\}$ are
uniquely identifiable. This shows that incorporating locally available entanglement in the probe design restores unique identifiability under imperfect \ac{BSMs} in the $n$-node star network.

In the proposed probe construction, the link and BSM parameters are identified jointly from the tomography probes without prior calibration of $b$. Within this joint-estimation setting, one additional independent probe relation is necessary and sufficient, and the $b$ versus $b^2$ dependence provides this relation. We do not claim that the construction is globally minimal in physical resources. Alternatively, a dedicated BSM-calibration experiment may be used to estimate $b$ separately, allowing the link parameters to be estimated independently.

\vspace{-1.0em}
\subsection{Fisher Information Matrix}
Having established the identifiability of the underlying parameters, we now quantify the accuracy with which they can be estimated. While identifiability guarantees that the parameters can be uniquely recovered, it does not characterize the achievable estimation precision. We therefore derive the \ac{FIM} associated with the measurements performed on the proposed probe states and use the corresponding \ac{CRB} to bound the estimation error.

The \ac{QFIM} generally contains contributions from derivatives of both the eigenvalues and eigenvectors of the density operator. For the Bell-diagonal probe states considered here, the eigenvectors are the parameter-independent Bell-basis states, so the eigenvector derivative term vanishes. Hence, the \ac{QFIM} equals the \ac{CFIM} of the eigenvalue distribution $\{\lambda_k\}$~\cite{liu2020quantum, optimalqnt}, and is given by
\begin{equation}
\label{11}
F_{ij} = \sum_{k} \frac{1}{\lambda_k}
\left( \frac{\partial \lambda_k}{\partial \theta_i} \right)
\left( \frac{\partial \lambda_k}{\partial \theta_j} \right).
\end{equation}
Since the corresponding \ac{SLDs} are diagonal in the same parameter-independent Bell basis, they commute, and the multiparameter \ac{QCRB} is attainable under Bell-basis measurements. 

For the single-link cyclic probe associated with $e_k$, the Bell-basis measurement is a measurement in the eigenbasis of the corresponding SLDs, so its \ac{QFI} is attained. From the outcome probabilities given in Eq.~\eqref{eq:single_link_prob}, the nonzero \ac{QFIM} entries are
\begin{align}
& F_{w_kw_k}^{(k)} = \frac{12w_k^2b^2}{(1+3w_k^2b)(1-w_k^2b)}, \\
& F_{bb}^{(k)} = \frac{3w_k^4}{(1+3w_k^2b)(1-w_k^2b)}, \text{ and }\\
& F_{w_kb}^{(k)} = \frac{6w_k^3b} {(1+3w_k^2b)(1-w_k^2b)}.
\end{align}
For the two-link probe associated with $e_u$ and $e_v$, the effective parameter is $z_{uv}=w_uw_vb^2$. The local $Z\otimes Z$ measurement distinguishes only the correlated outcomes $(00,11)$ from the anticorrelated outcomes $(01,10)$ with corresponding outcome probabilities given by
\begin{equation}
\label{eq:indirect_probs}
p_{00}=p_{11}
=
\frac{1+w_uw_vb^2}{4},
\qquad
p_{01}=p_{10}
=
\frac{1-w_uw_vb^2}{4},
\end{equation}
Therefore, the resulting entries correspond to the \ac{CFI} of the local measurement and are not generally equal to the \ac{QFIM} of the two-link probe state. The nonzero entries are
\begin{align}
F_{w_uw_u}^{(uv)}
&=
\frac{w_v^2b^4}
{1-w_u^2w_v^2b^4},
&
F_{w_vw_v}^{(uv)}
&=
\frac{w_u^2b^4}
{1-w_u^2w_v^2b^4},\\
F_{bb}^{(uv)}
&=
\frac{4w_u^2w_v^2b^2}
{1-w_u^2w_v^2b^4},
&
F_{w_uw_v}^{(uv)}
&=
\frac{w_uw_vb^4}
{1-w_u^2w_v^2b^4},\\
F_{w_ub}^{(uv)}
&=
\frac{2w_uw_v^2b^3}
{1-w_u^2w_v^2b^4},
&
F_{w_vb}^{(uv)}
&=
\frac{2w_u^2w_vb^3}
{1-w_u^2w_v^2b^4}.
\end{align}

Each probe contributes additively to the total \ac{FIM}. Since the probe set combines the \ac{QFI} attained by the single-link measurements with the \ac{CFI} of the two-link measurements, we refer to the combined matrix and its associated bound as the \ac{FIM} and \ac{CRB}, respectively. Hence, the overall FIM for the parameter vector $\boldsymbol{\theta}=(w_0,\ldots,w_{L - 1},b)$ is
\begin{equation}
\mathbf{F}(\boldsymbol{\theta})
=
\sum_p
\mathbf{F}^{(p)}(\boldsymbol{\theta}),
\end{equation}
where $\mathbf{F}^{(p)}$ denotes the contribution of the $p^{\mathrm{th}}$ probe embedded in the full parameter space. For $N$ independent repetitions of each probe measurement, $\mathbf{F}^{(N)}(\boldsymbol{\theta})
=
N\mathbf{F}(\boldsymbol{\theta}).$
Therefore, for any unbiased estimator $\hat{\boldsymbol{\theta}}$,
$
\mathrm{Cov}(\hat{\boldsymbol{\theta}})
\succeq
\left(\mathbf{F}^{(N)}(\boldsymbol{\theta})\right)^{-1}$,
and
$\left(\mathbf{F}^{(N)}(\boldsymbol{\theta})\right)^{-1}$ is equal to $\frac{1}{N}\mathbf{F}^{-1}(\boldsymbol{\theta})$.
Thus, provided that the total FIM is full rank, the achievable estimation variance decreases inversely with the number of measurement samples.
\vspace{-1.0em}
\subsection{Maximum Likelihood Estimation}
We estimate the unknown parameters
$(w_0,\ldots,w_{L - 1},b)$ using maximum likelihood estimation. Under standard regularity conditions, the MLE is asymptotically efficient, so its covariance approaches the CRB determined by the Fisher information of the chosen measurements. 

For the single-link probe associated with link $e_k$, let $N_t^k$ denote the total number of Bell-basis outcomes and let $N_{\Phi^+}^k$ denote the number of $\Phi^+$ outcomes. The remaining Bell-basis outcomes are represented as $N^k=N_t^k-N_{\Phi^+}^k$.
For the two-link probe involving links $e_u$ and $e_v$, we define $N_+^{uv}=N_{00}^{uv}+N_{11}^{uv}$ and
$N_-^{uv}=N_{01}^{uv}+N_{10}^{uv}$,
where $N_{00}^{uv}$, $N_{11}^{uv}$, $N_{01}^{uv}$, and $N_{10}^{uv}$ denote the number of corresponding local Z-basis measurement outcomes obtained from the two-link probe.
Using~\eqref{eq:single_link_prob}~and~\eqref{eq:indirect_probs}, the log-likelihood functions $\ell_D$ and $\ell_I$ for the single-link and two-link probes are given by
\begin{equation}
\label{s}
\ell_{\mathrm{D}}
=
\sum_{k=0}^{L - 1}
\left[
N_{\Phi^+}^k
\log\left(\frac{1+3w_k^2b}{4}\right)
+
N^k
\log\left(\frac{1-w_k^2b}{4}\right)
\right], 
\end{equation}
and
\begin{equation}
\label{t}
\ell_{\mathrm{I}}
=
N_+^{uv}
\log\left(\frac{1+w_uw_vb^2}{4}\right)
+
N_-^{uv}
\log\left(\frac{1-w_uw_vb^2}{4}\right),
\end{equation}
respectively.

The total log-likelihood is $
\ell(\boldsymbol{\theta})
=
\ell_{\mathrm{D}}+\ell_{\mathrm{I}}.
$
This likelihood is nonlinear in the physical parameters because the probe statistics depend on multiplicative combinations of $w_k$ and $b$. By the invariance property of \ac{MLEs} \cite{zehna1966invariance}, the effective parameters are first estimated directly from the measurement data, after which the corresponding inverse transformation is applied to recover the physical link and \ac{BSM} parameters. For the single-link probe, the closed-form MLE of
$x_k=w_k^2b$ is
\begin{equation}
\label{x:mle}
\hat{x}_k
=
\frac{4N_{\Phi^+}^k-N_t^k}{3N_t^k},
\qquad
k=0,\ldots,L - 1.
\end{equation}
For the two-link probe, the closed-form MLE of
$z_{uv}=w_uw_vb^2$ is
\begin{equation}
\label{z:mle}
\hat{z}_{uv}
=
\frac{N_+^{uv}-N_-^{uv}}
{N_+^{uv}+N_-^{uv}}.
\end{equation}
Since $z_{uv}
=
w_uw_vb^2
=
b\sqrt{x_ux_v}$,
the imperfect-\ac{BSM} parameter is estimated as
\begin{equation}
\label{b:mle}
\hat{b}
=
\frac{\hat{z}_{uv}}
{\sqrt{\hat{x}_u\hat{x}_v}}.
\end{equation}
The link parameters are then estimated as
\begin{equation}
\label{w:mle}
\hat{w}_k
=
\sqrt{\frac{\hat{x}_k}{\hat{b}}},
\qquad
k=0,\ldots,L - 1.
\end{equation}
Thus, single-link probes together with one two-link probe are sufficient to estimate all $L+1$ unknown parameters in the $n$-node star network. At finite $N$, sampling fluctuations may yield infeasible raw estimates, defined as cases in which the closed-form estimators in Eqs.~\eqref{x:mle}--\eqref{w:mle} are undefined or produce at least one parameter outside the physical region $0<\hat{w}_k,\hat{b}<1$. In such cases, we use a constrained MLE over $0<w_k,b<1$, and the probability of requiring constrained optimization decreases with $N$, as confirmed by the Monte Carlo analysis in the following section. Under standard regularity conditions, the MLEs are consistent and asymptotically efficient, although finite-sample bias may occur. Hence, as $N$ increases, their covariance approaches the CRB; for measurements that attain the QFI, the corresponding CRB coincides with the QCRB.

\section{Numerical Analysis}
\begin{figure}[t]
\vspace{-4mm}
\centering
\hspace{0.5mm}
\subfloat[\label{q22}]{\includegraphics[scale=0.25]{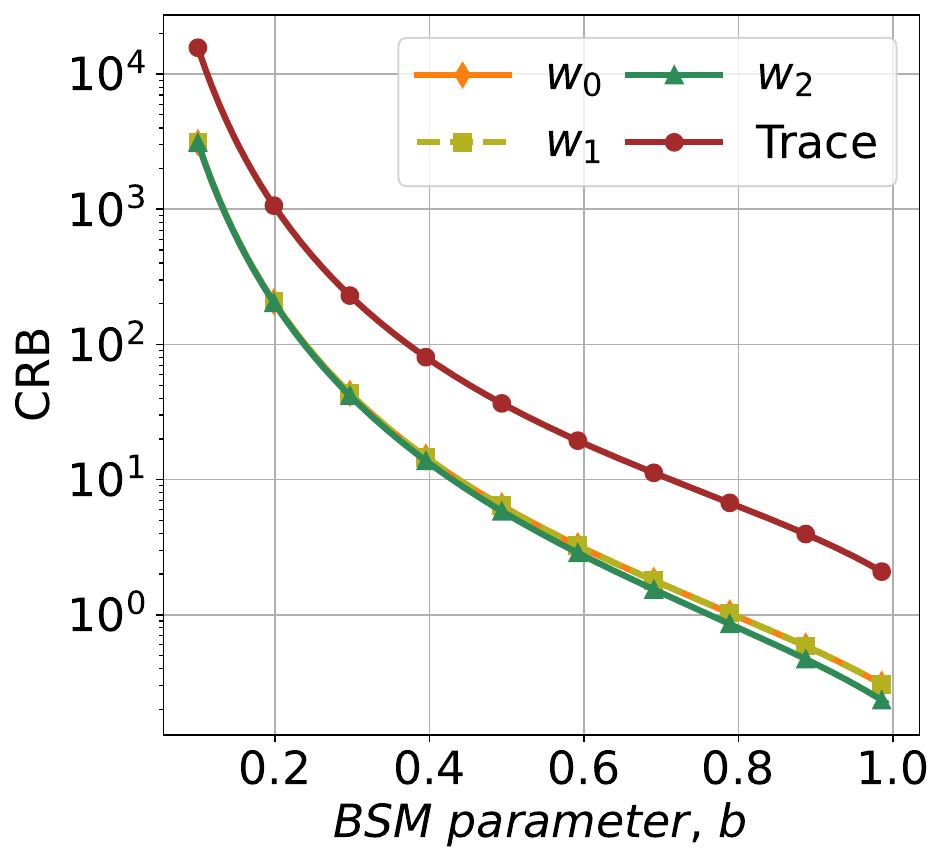}}%
\hfill
\subfloat[\label{q44}]{\includegraphics[scale=0.25]{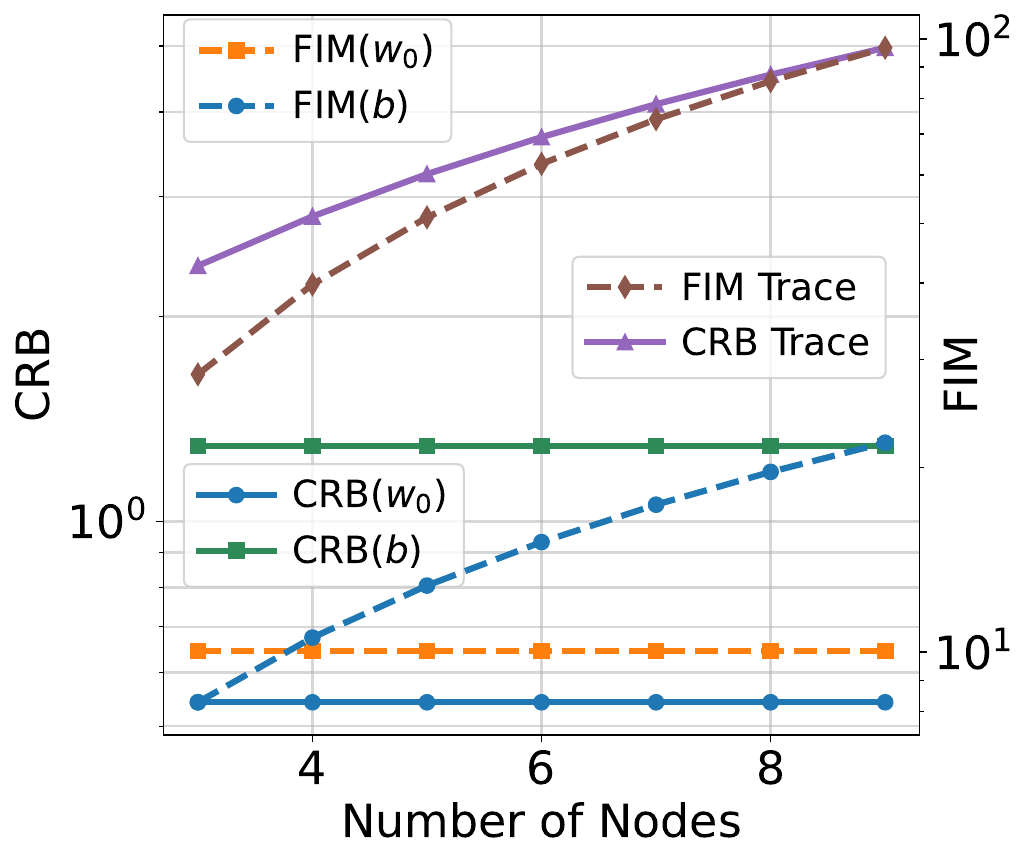}}%
\vspace{-0.5em}
\caption{FIM and CRB analysis under imperfect BSMs with link parameters $w_k=0.9$: (a) $[\mathbf{F}^{-1}]_{kk}$, for $k \in \{0,1,2\}$ and $\operatorname{Tr}(\mathbf{F}^{-1})$ versus $b$ for a four-node star network; (b) $F_{kk}$, $[\mathbf{F}^{-1}]_{kk}$ for $k=0$, together with $F_{bb}$, $[\mathbf{F}^{-1}]_{bb}$, $\operatorname{Tr}(\mathbf{F})$, and $\operatorname{Tr}(\mathbf{F}^{-1})$ versus number of nodes for $b=0.9$.}
\label{2}
\end{figure}

In this section, we present numerical evaluation results of the proposed probe design. Fig.~\ref{q22} shows the effect of the imperfect-\ac{BSM} parameter on the estimation precision in an four-node star network. For small $b$, the \ac{BSM} is highly noisy, making the effective probe parameters $w_k^2 b$ and $w_u w_v b^2$ weak and leading to large \ac{CRB} values. As $b$ increases, the \ac{BSM} approaches the ideal case, the probes become more informative, and the \ac{CRB} decreases. The parameters $w_0$ and $w_1$ show identical behavior due to symmetry under homogeneous noise, whereas the additional link parameter $w_2$ has a lower \ac{CRB} because it is estimated only through a single-link probe and is, therefore, less affected by the coupling introduced by the two-link probe. The aggregate \ac{CRB}, i.e., $\operatorname{Tr}(\mathbf{F}^{-1})$ follows the same decreasing trend but remains larger because it represents the total uncertainty over all estimated parameters. 

Fig.~\ref{q44} shows the variation of the \ac{FIM} and \ac{CRB} with network size under homogeneous noise, where the number of nodes varies from three to nine. As the network grows, the aggregate \ac{FIM}, i.e., $\operatorname{Tr}(\mathbf{F})$ increases because each added node introduces a new link and an additional cyclic probe of the form $\rho(w_k^2 b)$. The \ac{FIM} entry for $b$, i.e., $F_{bb}$ also increases since $b$ appears in all probes. However, the \ac{CRB} of $b$, i.e., $[\mathbf{F}^{-1}]_{bb}$ does not necessarily decrease, as each new probe also introduces an additional unknown link parameter coupled with $b$. The aggregate \ac{CRB} increases because the number of unknown parameters grows with the network size. The FIM and the CRB of link parameters, i.e., $F_{kk}$ and $[\mathbf{F}^{-1}]_{kk}$ remain nearly constant as the network size increases since each link parameter appears only in a fixed set of probes. 

\begin{figure}[t!]
\vspace{-4mm}
\centering
\subfloat[\label{m1}]{\includegraphics[scale=0.25]{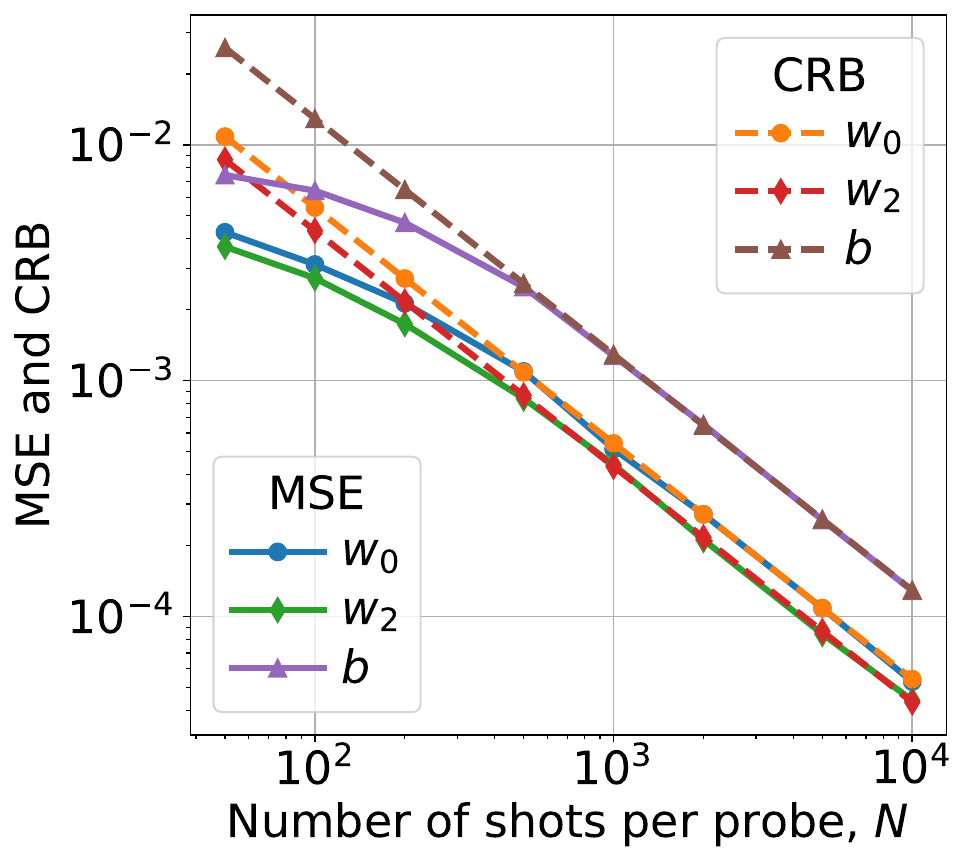}}%
\hfill
\subfloat[\label{m2}]{\includegraphics[scale=0.25]{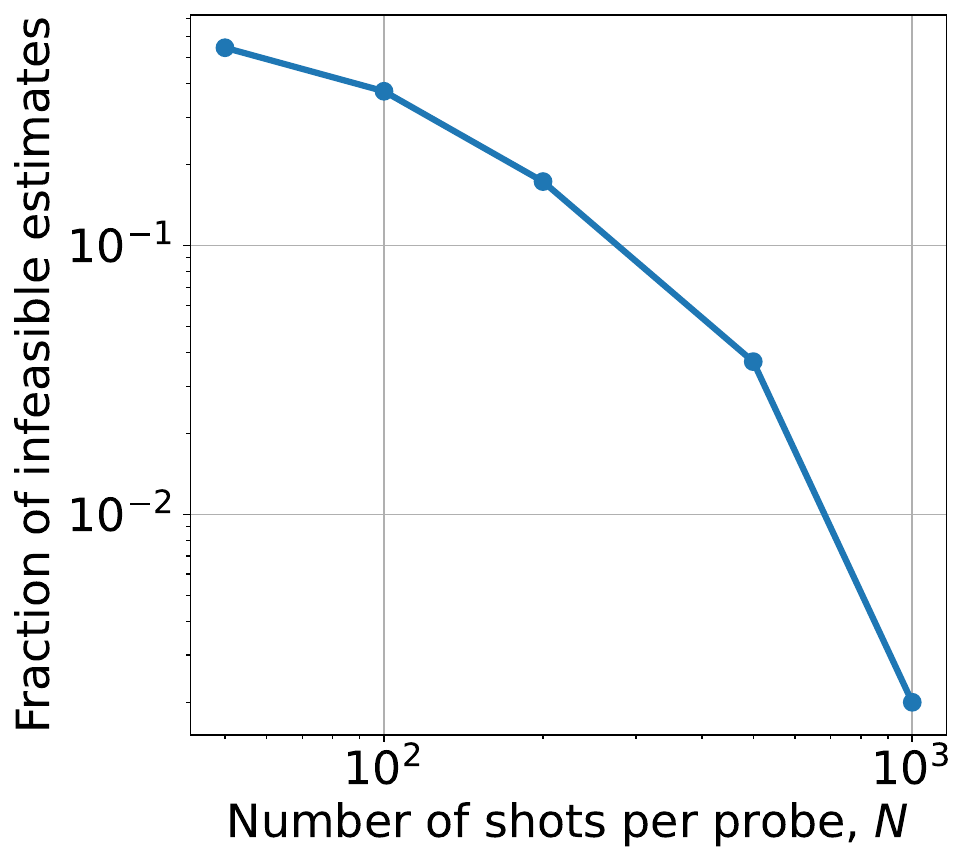}}%
\hspace*{2.5mm}
\vspace{-0.5em}
\caption{Monte Carlo analysis over $2000$ trials for a four-node star network with $w_0=w_1=w_2=0.9$ and $b=0.9$: (a) empirical MSE and corresponding CRB versus number of shots per probe; (b) fraction of infeasible estimates versus number of shots per probe.}
\label{3}
\end{figure}

We next evaluate the finite-sample performance of the proposed estimators through Monte Carlo simulation by comparing their empirical MSEs with the corresponding CRBs and examining the occurrence of infeasible estimates. As shown in Fig.~\ref{m1}, the empirical MSEs of $\hat{w}_0$, $\hat{w}_2$, and $\hat{b}$ decrease approximately as $1/N$ and approach the corresponding CRBs as $N$ increases, indicating the consistency and asymptotic efficiency of the estimators. At small $N$, the constrained estimator may be biased, and its MSE can therefore fall below the CRB for unbiased estimators. Fig.~\ref{m2} shows that the fraction of infeasible raw estimates decreases from approximately $0.55$ at $N=50$ to nearly zero at $N=1000$, confirming that the need for constrained likelihood optimization occurs mainly at finite sample sizes.

\vspace{-1.0em}
\section{Conclusion}
In this work, we investigated the problem of identifiability and estimation precision in quantum network tomography under imperfect \ac{BSMs}. We showed that, due to the multiplicative coupling between link and BSM depolarizing parameters, existing probe generation approaches are insufficient to uniquely identify individual parameters. To address this challenge, we introduced a locally generated entanglement resource that enables probe designs yielding independent parameter combinations that ensure unique identifiability of all parameters. In addition, we derived closed-form expressions for the \ac{FIM} and \ac{MLEs}, and evaluated the resulting estimation precision in an $n$-node star network based on the \ac{CRB}. Numerical results show that BSM imperfections increase estimation error, while the proposed probes maintain nearly stable precision for individual link parameters as the network scales, despite increasing aggregate uncertainty. Monte Carlo results further show that the MSE converges to the CRB as the number of measurements increases. Future work will extend the framework to general topologies and incorporate additional non-idealities such as memory decoherence and loss.
\vspace{-0.5em}
\section*{Acknowledgments}
This research was supported by SFI (Science Foundation Ireland) grant 21/US-C2C/3750 for CoQREATE (Convergent Quantum REsearch Alliance in Telecommunications), CONNECT-2 grant 13/RC/2077\_P2, and by the NSF-ERC Center for Quantum Networks grant EEC- 1941583.

\bibliographystyle{IEEEtran}
\bibliography{bibliography}

\vfill

\end{document}